 \documentclass[final,5p,times,twocolumn]{elsarticle}

\usepackage{amssymb}

\journal{Physics Letters B}
\newcommand{\nuc}[2]{\hbox{$^{#1}$#2}}
\begin{document}

\begin{frontmatter}



\title{In-beam $\gamma$-ray spectroscopy at the proton dripline: \nuc{40}{Sc}}


\author[NSCL,MSUPA]{A. Gade}
\author[NSCL]{D. Weisshaar}
\author[NSCL,MSUPA] {B. A. Brown}
\author[SUR]{J. A. Tostevin}
\author[NSCL,MSUPA]{D. Bazin}
\author[NSCL,MSUC] {K. Brown}
\author[WashU]{R. J. Charity}
\author[NSCL,MSUPA]{P. J. Farris}
\author[NSCL,MSUPA]{A. M. Hill}
\author[NSCL] {J. Li}
\author[NSCL,MSUPA]{B. Longfellow}
\author[ANL]{W. Reviol}
\author[NSCL,MSUPA] {D. Rhodes}

\address[NSCL]{National Superconducting Cyclotron Laboratory, East
Lansing, Michigan 48824, USA}
\address[MSUPA]{Department of Physics \& Astronomy, Michigan State
  University, East Lansing, Michigan 48824, USA}
\address[SUR] {Department of Physics, Faculty of Engineering and Physical Sciences, University of Surrey, Guildford, Surrey GU2 7XH, United Kingdom}
\address[MSUC]{Department of Chemistry, Michigan State
  University, East Lansing, Michigan 48824, USA}
\address[WashU]{Departments of Chemistry, Washington University, St. Louis, Missouri 63130, USA}
\address[ANL] {Physics Division, Argonne National Laboratory, Argonne, Illinois 60439, USA}

\begin{abstract}
We report on the first in-beam $\gamma$-ray spectroscopy of the proton-dripline nucleus
\nuc{40}{Sc} using  two-nucleon pickup onto an intermediate-energy rare-isotope beam of \nuc{38}{Ca}. The  \nuc{9}{Be}(\nuc{38}{Ca},\nuc{40}{Sc}$+\gamma$)X reaction at 60.9 MeV/nucleon mid-target energy selectively populates states in \nuc{40}{Sc} for which the transferred proton and neutron couple to high orbital angular momentum. In turn, due to angular-momentum selection rules in proton emission and the nuclear structure and energetics of \nuc{39}{Ca}, such states in \nuc{40}{Sc} then exhibit $\gamma$-decay branches although they are well above the proton separation energy.  This work uniquely complements results from particle spectroscopy following charge-exchange reactions on \nuc{40}{Ca} as well as \nuc{40}{Ti} EC/$\beta^+$ decay which both display very different selectivities. The population and $\gamma$-ray decay of the previously known first $(5^-)$ state at 892~keV and the observation of a new level at 2744~keV are discussed in comparison to the mirror nucleus and shell-model calculations. On the experimental side, this work shows that high-resolution in-beam  $\gamma$-ray spectroscopy is possible with new generation Ge arrays for reactions induced by rare-isotope beams on the level of a few $\mu$b of cross section.
\end{abstract}




\end{frontmatter}

\begin{figure}[h]
\includegraphics[width=0.45\textwidth]{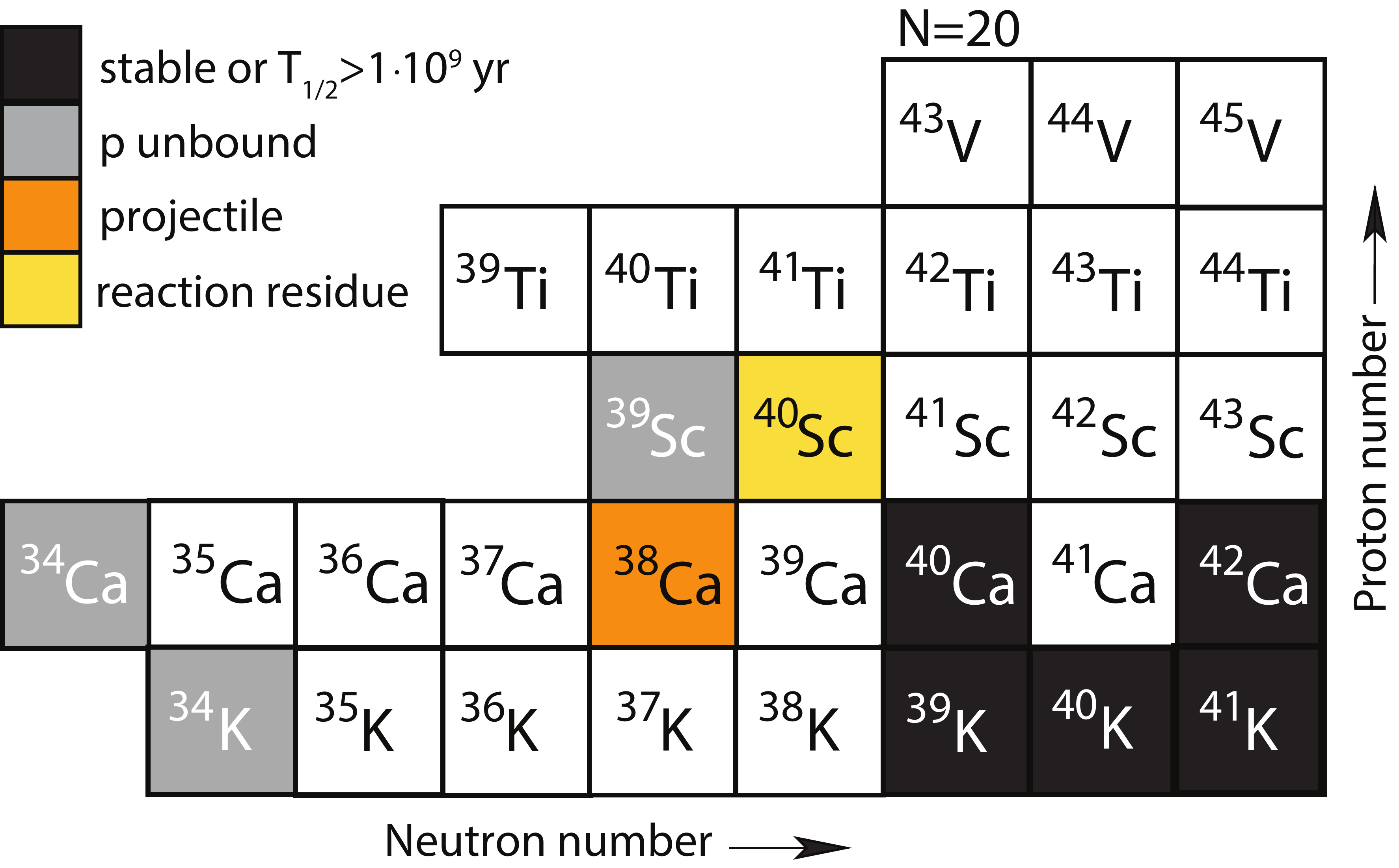}
\caption{\label{fig:chart} Part of the nuclear chart around \nuc{40}{Sc}. In fact, \nuc{40}{Sc} is the heaviest dripline nucleus for which the directly neighboring isobar (\nuc{40}{Ca}) is actually stable, allowing for extensive charge-exchange studies with stable beams and targets. The only other such isobar pair in the $sd$ shell or above is \nuc{20}{Na} (dripline) - \nuc{20}{Ne} (stable).  Nevertheless, $\gamma$-ray spectroscopy of \nuc{40}{Sc} had never been performed.}
\end{figure}

Since its discovery in 1955~\cite{gla55}, the neutron-deficient
nucleus \nuc{40}{Sc} has attracted attention for a variety of interests ranging from $rp$-process nucleosynthesis~\cite{han00,ill10}  to the solar neutrino absorption rate on \nuc{40}{Ar}~\cite{orm95,bha98}. In fact, \nuc{40}{Sc} -- five neutrons removed from stable \nuc{45}{Sc} -- is the last proton-bound scandium isotope, with \nuc{39}{Sc} shown to be unstable against proton emission~\cite{woo88}. \nuc{40}{Sc} is peculiarly located on the nuclear chart (Fig.~\ref{fig:chart}): While it is the proton dripline nucleus of the scandium isotopic chain, it is easily produced from charge-exchange reactions on stable \nuc{40}{Ca} (e.g., see \cite{han00,chi86,tab84}).

Due to the low \nuc{40}{Sc} proton separation energy of $S_p=529.6(29)$~keV~\cite{ame16},  only the $4^-$ ground state and the 34-keV first-excited $(3^-)$ state are nominally below the proton emission threshold. The nuclear structure interest in this neighboring isobar of \nuc{40}{Ca} has been focused on the particle-hole nature of the states in \nuc{40}{Sc} relative to the doubly-magic $N=Z=20$ core~\cite{chi86,loi71}, while the quest to constrain the \nuc{39}{Ca}($p,\gamma$)\nuc{40}{Sc} proton capture rate drove the highest-resolution study of  \nuc{40}{Sc} yet~\cite{han00}. To obtain the \nuc{40}{Ti}$\rightarrow$\nuc{40}{Sc} weak decay rate, which allows determination of  the \nuc{40}{Ar} neutrino absorption rate via isospin symmetry~\cite{orm95}, the $\beta$ decay of \nuc{40}{Ti}, populating high-lying, unbound low-spin states of \nuc{40}{Sc}, was studied with proton spectroscopy (e.g., see \cite{bha98,liu98}). The work reported here presents the first in-beam $\gamma$-ray spectroscopy of this dripline nucleus, \nuc{9}{Be}(\nuc{38}{Ca},\nuc{40}{Sc}$+\gamma$)X, including observation of decays from states above $S_p$.

\begin{figure}[h]
\includegraphics[width=0.45\textwidth]{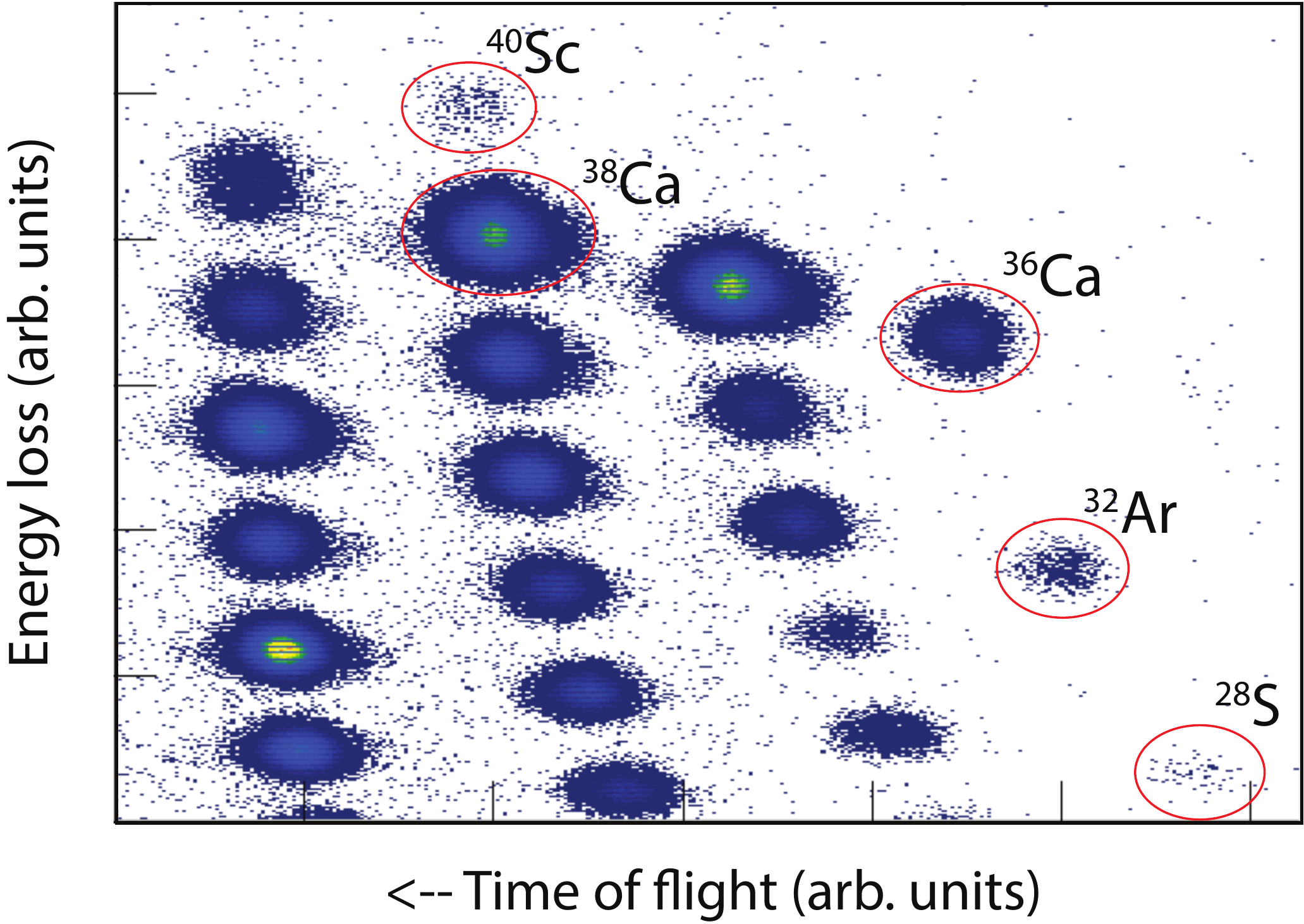}
\caption{\label{fig:pid} Event-by-event particle identification, energy loss vs. time of flight, of the reaction residues produced in \nuc{38}{Ca} + \nuc{9}{Be} at 61~MeV/nucleon (mid-target). The energy loss was measured with the S800 ionization chamber and the time of flight was taken between two plastic scintillators in the S800 analysis beam line and at the back of the S800 focal plane. To show the reaction residues together with a tail of the \nuc{38}{Ca} projectiles entering the focal plane, a particle-$\gamma$ coincidence trigger was required for the purpose of the figure. A number of (near) dripline reaction residues are marked (the data runs used for the cross section determination are displayed). }
\end{figure}

The \nuc{38}{Ca} secondary beam was produced by fragmentation
of a 140-MeV/nucleon stable \nuc{40}{Ca} beam, accelerated
by the Coupled Cyclotron Facility at NSCL~\cite{gad16}, impinging
on a 799 mg/cm$^2$ \nuc{9}{Be} production target and separated
using a 300 mg/cm$^2$ Al degrader in the A1900 fragment
separator~\cite{a1900}. The momentum acceptance of the separator
was restricted to $\Delta p/p=0.25$\%, yielding typical rates of
160,000 \nuc{38}{Ca}/s. About 86\% of the secondary beam composition
was \nuc{38}{Ca}, with the lighter isotones comprising the less intense beam components.
The secondary \nuc{9}{Be} reaction target, of 188 mg/cm$^2$ thickness,
was located at the target position of the S800 spectrograph.
The projectile-like reaction products were identified on an event-by-event
basis in the S800 focal plane with the standard detector systems \cite{s800} (see Fig.~\ref{fig:pid}). The \nuc{38}{Ca} projectiles in the entrance channel were selected through a software gate applied on the time-of-flight difference taken between two plastic scintillators before the target.

\begin{figure}[h]
\includegraphics[width=0.45\textwidth]{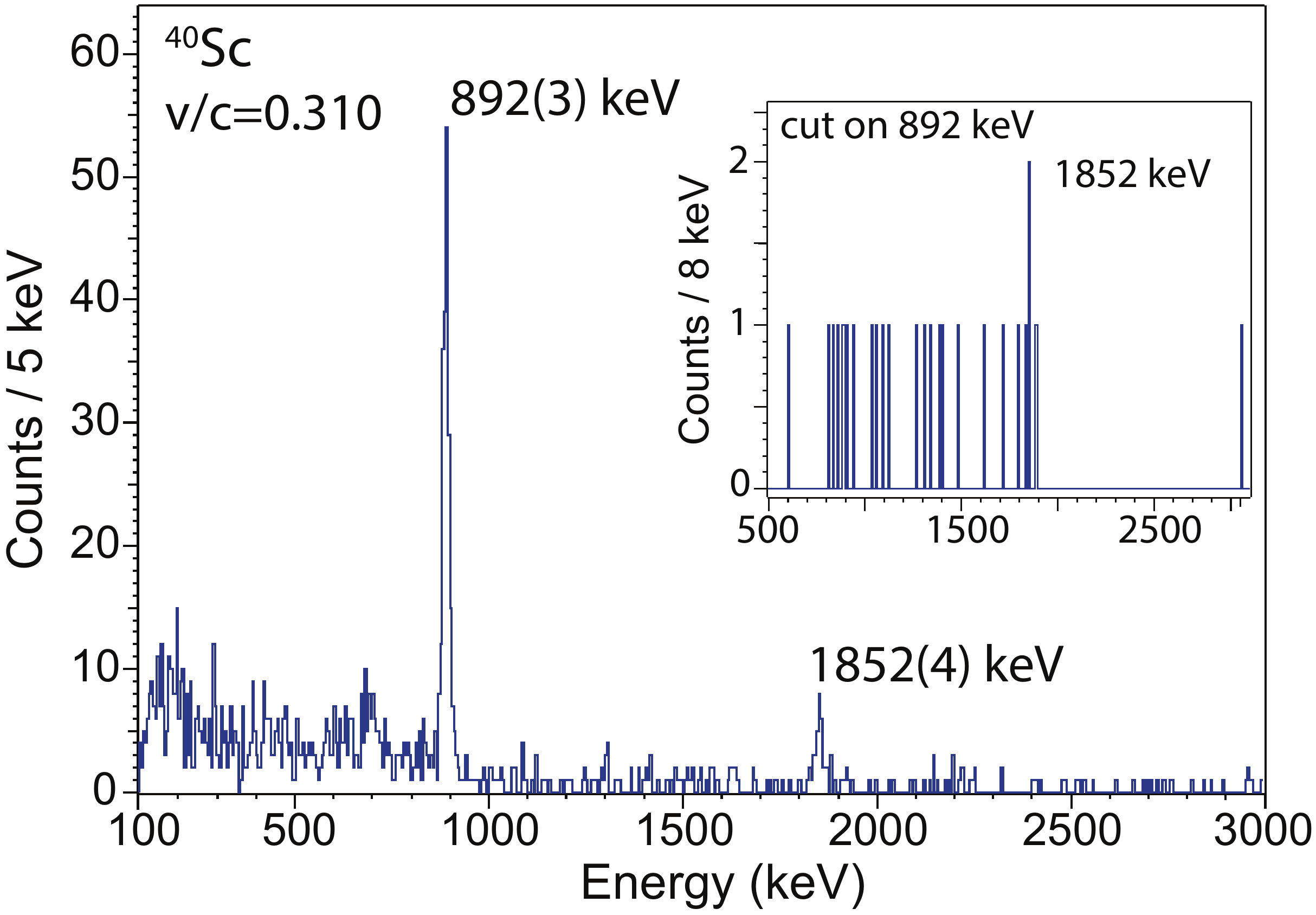}
\caption{\label{fig:gamma} Doppler-reconstructed $\gamma$-ray
spectrum detected with GRETINA in coincidence with \nuc{40}{Sc} reaction residues
produced in the two-nucleon pickup onto \nuc{38}{Ca}. The 892-keV $\gamma$-ray transition corresponds to the de-excitation of the known $(5^-)$ state reported at 893.5 keV~\cite{nndc} to the ground state. The second $\gamma$-ray cannot be attributed to an already known state in \nuc{40}{Sc}. The inset shows the $\gamma$-ray spectrum in coincidence to the 892-keV transition. Despite the very low statistics, the spectrum is consistent with 1852 and 892~keV forming a cascade.}
\end{figure}

The high-resolution $\gamma$-ray spectrometer GRETINA
\cite{pas13,wei17}, an array of 36-fold segmented high-purity germanium
detectors assembled into modules of four crystals each, was
used to measure the prompt $\gamma$ rays emitted by the reaction
residues in flight. The 12 detector modules available were
arranged in two rings with four located at 58$^\circ$ and eight at 90$^\circ$
with respect to the beam axis. Online pulse-shape analysis
provided the $\gamma$-ray interaction points for event-by-event Doppler
reconstruction of the $\gamma$ rays emitted in-flight  at about 30\% of the speed of light~\cite{wei17}. The momentum vector of projectile-like reaction residues as ray-traced through the S800 spectrograph
was incorporated into the emission-angle determination entering Doppler reconstruction. Figure~\ref{fig:gamma}
displays the Doppler-reconstructed $\gamma$-ray spectrum obtained for \nuc{40}{Sc} with nearest-neighbor addback included \cite{wei17}.

\begin{figure}[h]
\includegraphics[width=0.4\textwidth]{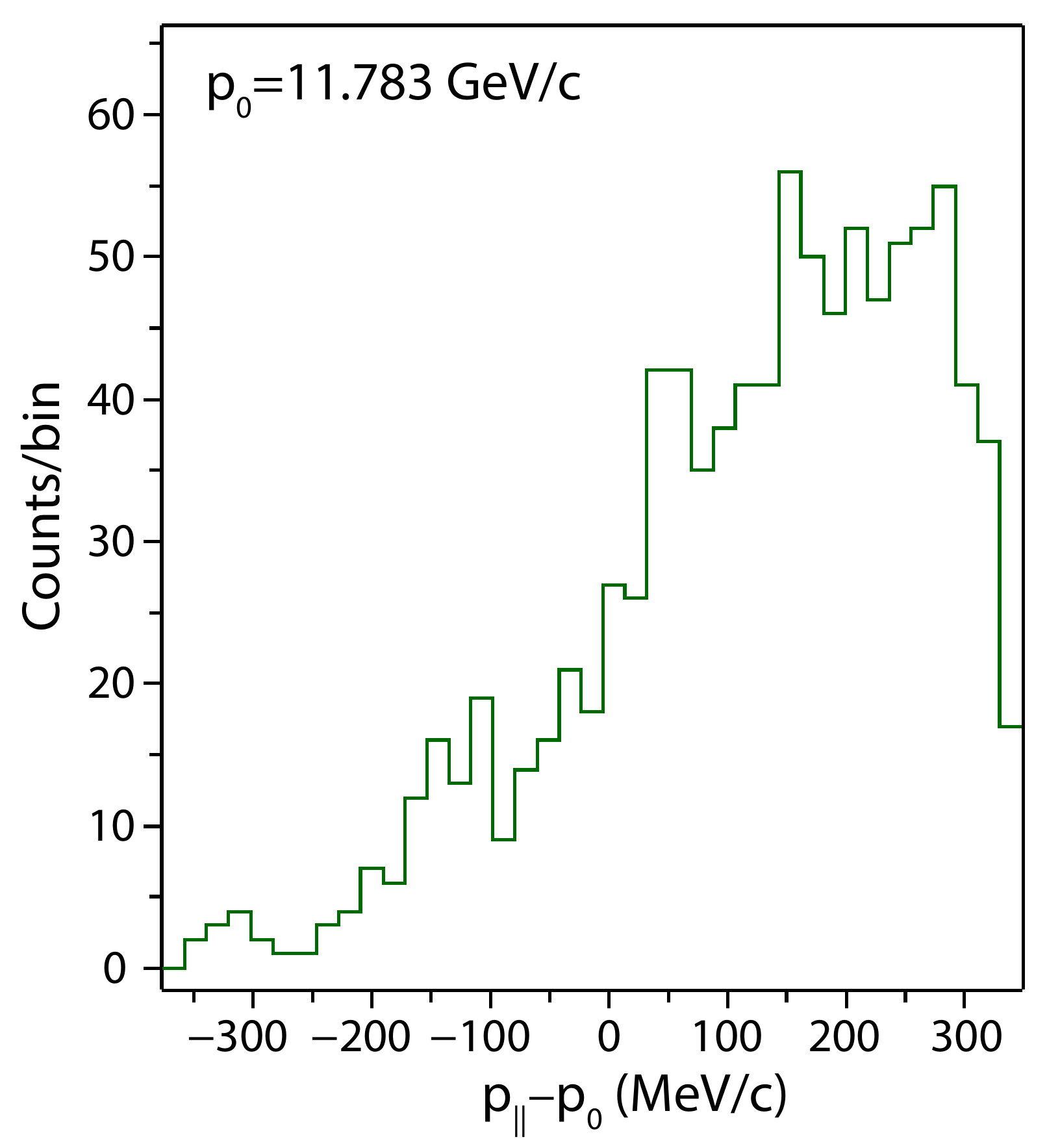}
\caption{\label{fig:ppar} Parallel momentum distribution of the \nuc{40}{Sc} reaction residues relative to the set value of the S800 spectrograph. The range shown corresponds to the nominal acceptance of its focal plane. The magnetic rigidity was set to center \nuc{36}{Ca}, placing the distribution of \nuc{40}{Sc} slightly towards the edge of the acceptance with potential losses. The shape of the distribution is reminiscent of the observations for the corresponding fast-beam one-nucleon pickup reactions explored earlier~\cite{gad08,gad11,gad07}. }
\end{figure}

The inclusive cross section for the
two-nucleon pickup from \nuc{38}{Ca} to \nuc{40}{Sc} was determined from the number of \nuc{40}{Sc} detected in the S800 focal plane relative to the number of \nuc{38}{Ca} projectiles and the number density of the target. The rigidity of the spectrograph was chosen to center the two-neutron knockout residue \nuc{36}{Ca} in the S800 focal plane and, therefore, \nuc{40}{Sc} was off-center. Figure~\ref{fig:ppar} shows the parallel momentum distribution of \nuc{40}{Sc} within the acceptance of the spectrograph. Assuming that the maximum of the distribution is at about 11.983~GeV/c (see Fig.~\ref{fig:ppar}) and has a shape similar to what was observed in~\cite{gad08} for one-proton pickup from a \nuc{9}{Be} target, a potential  acceptance loss of 20\% is estimated~\footnote{We note that the exact shape and centroid of the momentum distribution from this novel \nuc{9}{Be}-induced reaction is not precisely known and future measurements of the shape and energetics may clarify the reaction mechanism and allow for a more precise estimate of the acceptance loss. This is not critical for the results of the present work.}.  Including this uncertainty, the inclusive cross section amounts to $\sigma_{inc}=8.0(6)^{+1.6}$~$\mu$b (with 3.75\% statistical and 7\% systematic uncertainty included in the symmetric error bars and additional +20\% of uncertainty accounting for a possible acceptance cut.). The systematic uncertainty is attributed to the determination of a very low cross section in the presence of background from pile-up. 

While, due to its unbound target final states, the present reaction mechanism 
is too complex to allow quantitative dynamical calculations, in common with 
other linear- and angular-momentum mismatched two-nucleon transfer reactions, 
such as ($\alpha,d$) and its inverse, see e.g. \cite{fis94,wuo14}, its strong 
selectivity of (stretched) transitions involving maximal orbital angular momentum 
transfer is a firm qualitative feature. Such large $\ell$-selectivity in one-neutron 
pickup at intermediate energy is shown in Fig. 2 of Ref. \cite{gad16p} and where, 
for a $^9$Be target, the reaction proceeds by the pickup of well-bound nucleons
leaving the target residue in the continuum \cite{gad11}. Importantly, unlike 
the ($\alpha,d$) reaction, where the transfer vertex selects an np-pair with
spin $S=1$, here there is no such restriction, allowing, for example, for the 
direct population of the $(\pi f_{7/2},\nu f_{7/2})^{(J=6^+)}$ final state. This 
difference is illustrated by the \nuc{38}{Ar}($\alpha,d$)\nuc{40}{K} reaction to 
the mirror of \nuc{40}{Sc} that was found to populate the $(\pi f_{7/2},\nu 
f_{7/2})^{(J=7^+)}$ configuration but not the corresponding $6^+$ state~\cite{del76} or by the  \nuc{40}{Ca}($\alpha,d$)\nuc{42}{Sc} reaction to the neighboring Sc isotope that populated the $7^+$ and $5^+$ states but not the $6^+$~\cite{nan77}. 

\begin{figure}[h]
\includegraphics[width=0.47\textwidth]{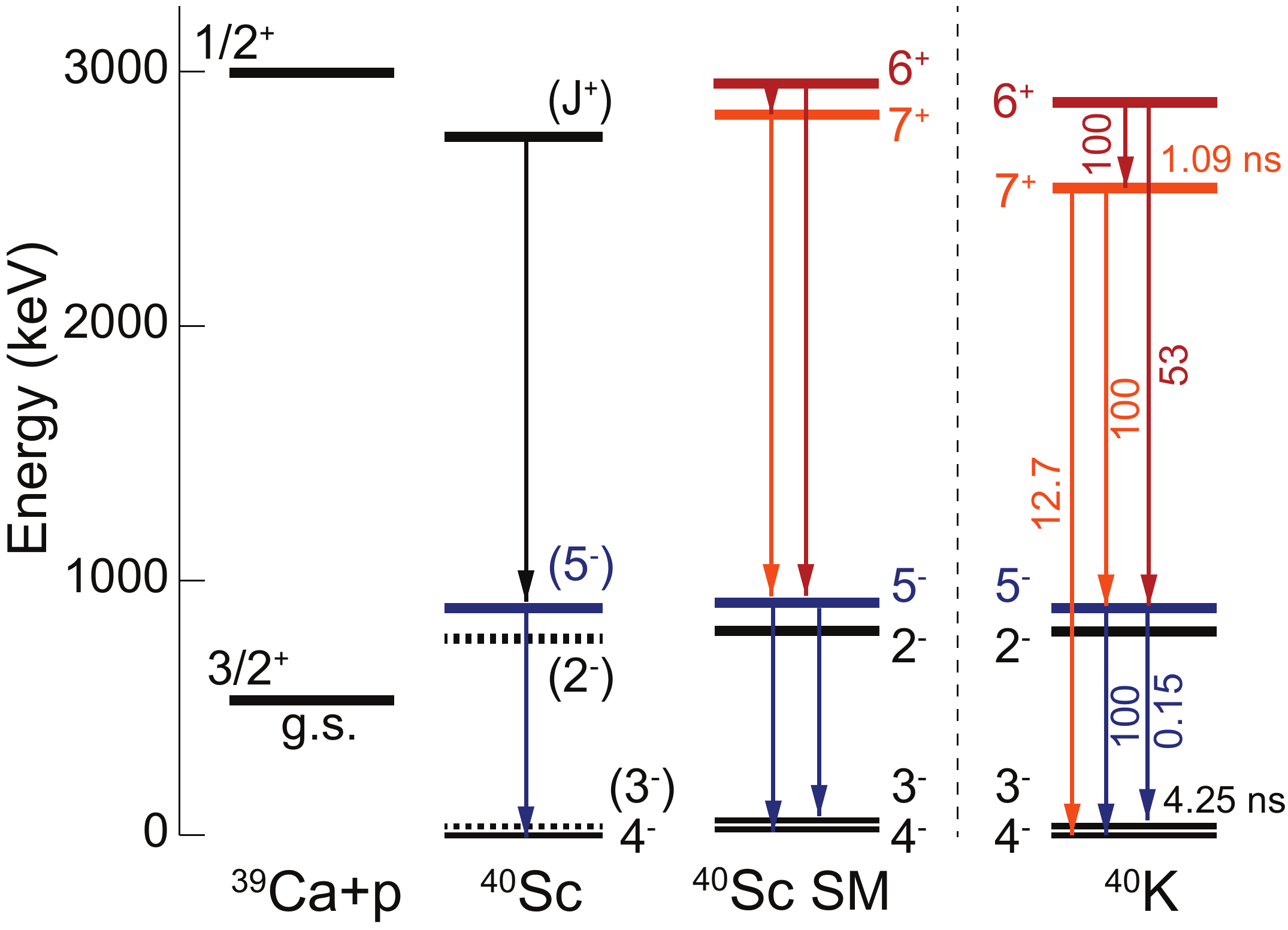}
\caption{\label{fig:level} Level schemes of the mirror pair \nuc{40}{Sc} and \nuc{40}{K} together with shell model for \nuc{40}{Sc} (using the $sdpf$-$wb$ Hamiltonian~\cite{war90}) and the \nuc{39}{Ca}$+p$ system relevant to explore proton emission from the relevant excited states in \nuc{40}{Sc}. For all states of \nuc{40}{Sc} discussed here, $p$ emission can only reach the $3/2^+$ ground state of \nuc{39}{Ca} due to the energetics of the two systems. Levels known in \nuc{40}{Sc} but not observed here are indicated by a dashed line. Literature data taken from~\cite{nndc}.}
\end{figure}

Turning to the $\gamma$-ray spectrum and the level structure of \nuc{40}{Sc}, the very favorable peak-to-background ratio manifested in Fig.~\ref{fig:gamma} enables the spectroscopy of rare isotopes produced at the level of $\mu$b. The $\gamma$ ray observed at 892(3)~keV (see Fig.~\ref{fig:gamma}) most certainly corresponds to the decay of the previously reported $(5^-)$ state at  893.5(20)~keV to the $4^-$ ground state~\cite{nndc}. Since this is the first $\gamma$-ray spectroscopy of \nuc{40}{Sc}, we resort to the mirror nucleus \nuc{40}{K} and shell-model calculations for guidance on other potential decay branches from this state. The shell model for \nuc{40}{Sc} uses the $sdpf$-$wb$ effective shell-model interaction~\cite{war90}, a $(sd)^{-1} (fp)^{+1}$ model space for the low-lying negative-parity states, and a $(sd)^{-2} (fp)^{+2}$ model space for the positive-parity states. In \nuc{40}{K}, the $5^- \rightarrow 4^-$ transition to the ground state dominates over the decay to the excited $3^-$ state with a branching ratio of 100 vs. 0.15 (see Fig.~\ref{fig:level}), consistent with the observation of only the 892~keV $\gamma$ ray here. This is also in agreement with the shell-model calculations that predict the $5^- \rightarrow 3^-$ branch is even more suppressed.

The population of the $5^-$ state in the reaction used here very likely corresponds to the pickup of the proton into the $f_{7/2}$ orbital and the neutron into the partially filled $d_{3/2}$ orbital, consistent with a resulting stretched configuration of $(\pi f_{7/2}^{+1},\nu d_{3/2}^{-1})^{(J=5^-)}$. The selectivity of the reaction mechanism favors population of high-orbital-angular-momentum states and, thus, supports this picture. The proton decay of the state is presumably hindered by the angular momentum barrier ($\ell=3$) and the low $Q_p$ value for the $p$ emission to the only energetically allowed state in \nuc{39}{Ca}, the $3/2^+$ ground state (see Fig.~\ref{fig:level}). The $4^-$ and $(3^-)$ ground and first-excited state are proposed to have the same $\pi f_{7/2} \nu d_{3/2}$ particle-hole configuration based on $(p,n)$ reaction studies~\cite{chi86} but their population would not be observable through prompt $\gamma$-ray spectroscopy (from the mirror nucleus, the $3^-$ state is expected to be a nanosecond isomer, also with the $\gamma$-ray energy below threshold in this work). The reaction mechanism also disfavors population of a $3^-$ configuration due to the lower orbital angular momentum transfer relative to the $5^-$ level.

In the following, we explore the origin of the $\gamma$-ray transition at 1852~keV. The next configuration that allows for high angular momentum can be realized by the pickup of the proton and neutron into the corresponding $f_{7/2}$ orbitals; our selectivity to high-angular-momentum configurations is again commensurate with the observation of a $\gamma$-ray decay. The highest $J^{\pi}$ states of the resulting $(f_{7/2})^2$ multiplet would be $6^+$ and $7^+$. In \nuc{40}{K}, the lowest-lying $7^+$ and $6^+$  states are reported at about 2.54 and 2.88~MeV excitation energy, respectively, both with decays to the $5^-$ state and to each other (Fig.~\ref{fig:level}). For \nuc{40}{Sc}, if the 1852-keV $\gamma$ ray, observed here for the first time, were to feed the $(5^-)$ state, this would place a new excited state at 2744(5)~keV in the region where the high-spin positive-parity states are expected. Also, the shell-model calculations performed using the $sdpf$-$wb$ Hamiltonian~\cite{war90} place these high-spin positive parity states in the same energy region (see Fig.~\ref{fig:level}). Turning to the mirror first, the $7^+$ state in \nuc{40}{K} is a nanosecond isomer due to the high $\gamma$-ray multipolarities involved (see Fig.~\ref{fig:level}). In weakly-bound \nuc{40}{Sc}, the $7^+$ state would be more than 2~MeV above the proton emission threshold, with the $\gamma$ decay hindered. Both the $6^+$ and $7^+$ states can decay by $\ell=4$ proton emission to the ground state of \nuc{39}{Ca}. For $Q_p=2.215$~MeV, the single proton decay width is 49~eV. The  $6^+$ $\gamma$ decay width is estimated to be 0.0020~eV (uncertain by up to a factor of 10) and, therefore, for the $6^+$ state to decay by $\gamma$-ray emission rather than proton decay, the $\pi g_{9/2}$ spectroscopic factor has to be of order $10^{-5}$, which is plausible but cannot be quantified with present shell-model Hamiltonians. The $\gamma$ width of the $7^+$, however, is smaller than that of the $6^+$ by about a factor of $10^{4}$, indicating that the $7^+$ level will likely decay by fast proton emission, given a $g_{9/2}$ spectroscopic factor of the order mentioned above, and would escape detection in the present experiment. Proton spectroscopy of these two states would indeed be interesting as the $\gamma$-$p$ competition provides information on the $g_{9/2}$ intrusion into the model spaces in this region which is otherwise out of reach.

Connecting this back to the reaction mechanism of two-nucleon pickup onto \nuc{38}{Ca}, the shell-model occupancies and two-nucleon amplitudes (TNAs) for \nuc{38}{Ca} and \nuc{40}{Sc} offer perspective. In terms of the [$(\pi d_{3/2}), (\pi f_{7/2}), (\nu d_{3/2}), (\nu f_{7/2})$] orbital occupancies, the dominant configuration of the $6^+$ and $7^+$ states in \nuc{40}{Sc} is
[4,1,2,1] (34\% and 65\%, respectively). The $0^+$  ground state of \nuc{38}{Ca} is dominated by [4,0,2,0] on the other hand. Thus, these two states under discussion are indeed populated by the addition of a proton and neutron into the $f_{7/2}$ orbitals, favored by the reaction mechanism used here.

So, $6^+$ remains as the likely assignment of the new state observed in \nuc{40}{Sc} but with the caveat that a strong $6^+ \rightarrow 7^+$ $\gamma$ branch would be expected based on the decay pattern of \nuc{40}{K}.  Using the branching ratio from \nuc{40}{K} and the intensity of the 1852-keV transition, about 185 counts would be expected at about 200~keV for a $6^+ \rightarrow 7^+$ transition based on the mirror. There is no evidence for such a strong transition anywhere in the spectrum (see Fig.~\ref{fig:gamma}).

The shell-model calculation, with a calculated $6^+$-$7^+$ energy spacing of only 127~keV for \nuc{40}{K} and \nuc{40}{Sc}, has the $6^+ \rightarrow 5^-$ branch as the strongest transition with $6^+ \rightarrow 7^+$ predicted to be only 1.5\% of that.  Adjusting the shell-model calculation so that it  modifies the $6^+$-$7^+$ energy gap to match the 336~keV observed in \nuc{40}{K} increases the $6^+ \rightarrow 7^+$  branch to 21\% relative to the strongest decay (to the $5^-$). The calculation with the $sdpf$-$wb$ Hamiltonian, which does not contain the Coulomb interaction, gives a similar result for \nuc{40}{K} and \nuc{40}{Sc}. The addition of the Coulomb interaction would change the mirror branching ratios in two ways. First, the $6^+ \rightarrow 5^-$ $B(E1)$ value could exhibit a mirror asymmetry. There are examples in this mass region where the mirror $B(E1)$ values differ by up to factors of ten~\cite{pat08}. Second, the $6^+$-$7^+$ spacing could change. For the dominant configurations of [4,1,2,1] for \nuc{40}{Sc} and [2,1,4,1] for \nuc{40}{K} the $6^+$-$7^+$ spacing is the same since the $f_{7/2}$ configuration is the same for both. The next most important configuration for the $6^+$ states is [3,2,3,0] for \nuc{40}{Sc} and [3,0,3,2] for \nuc{40}{K}. From experiment, the $6^+$ member of the proton $(f_{7/2})^2$ configuration in \nuc{42}{Ti} is lowered by 149~keV compared to the neutron $(f_{7/2})^2$ configuration in \nuc{42}{Ca} (see Fig. 1 in Ref.~\cite{kut78}).
Such a shift lowers the $6^+$ state in \nuc{40}{Sc} by 76~keV compared to \nuc{40}{K}  and reduces the branching to the $7^+$ to 11\% relative to  the $6^+ \rightarrow 5^-$  branch. Assuming a $6^+ \rightarrow 7^+$ branching of 21\% relative to the $6^+ \rightarrow 5^-$ transition would lead to about 20 counts expected in the low-energy region of the $\gamma$-ray spectrum (see Fig.~\ref{fig:gamma}). We do not see evidence in the spectrum but cannot exclude it either at the present level of statistics. This makes the data compatible with a scenario close to the shell-model calculations but would require  the aforementioned mirror asymmetry in the $6^+ \rightarrow  5^-$ $E1$ decay to explain the mirror difference in the branching ratio of the $6^+$ state between \nuc{40}{K} and \nuc{40}{Sc}.

Assuming the placement of the $\gamma$-ray transitions in \nuc{40}{Sc} as proposed in Fig.~\ref{fig:level} and supported by the low-statistics coincidence of Fig.~\ref{fig:gamma}, 58(8)\% of the cross section feeds the $(5^-)$ state at 892~keV and 22\% the $(J^+)$ level at 2744 keV. This leaves 20(2)\% of the inclusive cross section not resulting in prompt or sufficiently strong $\gamma$ rays. Consequently, this is the amount of cross section that could be carried by the  $4^-$ ground state and the potential $(3^-)$ nanosecond isomer.

In summary, we report the first $\gamma$-ray spectroscopy of the proton dripline nucleus \nuc{40}{Sc}, using a two-nucleon pickup reaction onto a fast rare-isotope beam of \nuc{38}{Ca}. Two excited states were observed to be populated, the previously known $(5^-)$ state at 892~keV and a new level proposed at 2744(5)~keV.  The nature of the states is discussed in comparison to the mirror nucleus \nuc{40}{K} and aided by the strong high-angular-momentum selectivity of the fast-beam pickup reaction. More broadly, this work demonstrates that in-beam $\gamma$-ray spectroscopy is possible with high-resolution enabled by new-generation germanium detection arrays on the level of a few $\mu$b of cross section. This work also marks the first exploration of such a fast-beam two-nucleon pickup reaction and consistency with the dominant role of momentum matching is shown as might have been expected from similar work on fast-beam one-nucleon pickup reactions.

This work was supported by the U.S. National Science Foundation (NSF) under Grant No. PHY-1565546, by the DOE National Nuclear Security Administration through the Nuclear Science and Security Consortium, under Award No. DE-NA0003180, and by the U.S. Department of Energy, Office of Science, Office of Nuclear Physics, under Grants No. DE-SC0020451 (MSU) and DE-FG02-87ER-40316 (WashU) and under Contract No. DE-AC02-06CH11357 (ANL). GRETINA was funded by the DOE, Office of Science. Operation of the array at NSCL was supported by the DOE under Grant No. DE-SC0019034. J.A.T acknowledges support from the Science and Technology Facilities Council (U.K.) Grant No. ST/L005743/1. B.A.B. acknowledges support from NSF Grant No. PHY-1811855.

\end{document}